\documentclass[aps,amsart]{article}
\usepackage{authblk}
\usepackage{xcolor}
\usepackage{changes}
\usepackage{tabularx}
 \usepackage{graphicx}
\usepackage{siunitx}
\usepackage{amsmath}
\usepackage{miller}
\usepackage{bm}
\usepackage{graphicx}
\usepackage{dcolumn}
\usepackage{float}
\usepackage{bm}

\begin{document}

\title{Small-angle neutron scattering of long-wavelength magnetic modulations in reduced sample dimensions}

\author[1\footnote{grace.causer@tum.de}]{Grace L. Causer}
\author[1]{Alfonso Chacon}
\author[2]{Andr{\'e} Heinemann}
\author[1,3,4]{Christian Pfleiderer}
\affil[1]{Physik-Department, Technical University of Munich, James-Franck-Stra{\ss}e 1, D-85748 Garching, Germany} 
\affil[2]{Heinz Maier-Leibnitz Zentrum (MLZ), Technical University of Munich, Garching, Germany} \affil[3]{Centre for QuantumEngineering (ZQE), Technical University of Munich, D-85748 Garching, Germany} 
\affil[4]{Munich Center for Quantum Science and Technology (MCQST), Technical University of Munich, D-85748 Garching, Germany.}

 \maketitle
\begin{abstract}

Magnetic small-angle neutron scattering (SANS) is ideally suited to provide direct, reciprocal-space information of long-wavelength magnetic modulations, such as helicoids, solitons, merons, or skyrmions. SANS of such structures in thin films or micro-structured bulk materials is strongly limited by the tiny scattering volume vis a vis the prohibitively large background scattering by the substrate and support structures. Considering near-surface scattering closely above the critical angle of reflection, where unwanted signal contributions due to substrate or support structures becomes very small, we establish equivalent scattering patterns of the helical, conical, skyrmion lattice, and fluctuation-disordered phases in a polished bulk sample of MnSi between conventional transmission and near-surface SANS geometries. This motivates the prediction of a complete repository of scattering patterns expected for thin films in the near-surface SANS geometry for each orientation of the magnetic order with respect to the scattering plane.

\end{abstract}

\maketitle

\pagebreak
\section{Introduction}

Since the discovery of topologically protected states in chiral magnets, the field of non-collinear magnetism has undergone a renaissance of research activity in recent years. This has led to the identification of many forms of non-collinear spin order in bulk materials, including skyrmions and anti-skyrmions \cite{science2009}, merons and anti-merons \cite{merons}, solitons \cite{PhysRevB.104.174420}, hopfions and chiral bobbers \cite{2019_Rybakov_PhysRevBb,doi:10.1073/pnas.1716887115,2018_Zheng_NatureNanotechc}. Essentially all of these forms of magnetic order are either the result of hierarchical coupling terms, or competing interactions generating various forms of frustration, both leading to the formation of nanoscale spin structures in real-space.

The associated topological properties of the spin structures enables several unusual properties, including highly efficient coupling to spin currents and an inherent stability even for metastable configurations \cite{2013_Nagaosa_NatureNanotechd, 2018_Bauer_TopologyinMagnetisma}. As a result, their technological relevance is widely discussed in the context of low-energy information carriers and data processing \cite{2020_Back_JPhysDApplPhysa,2020_Zhang_JPhysCondensMatter}. Here, the drive towards device miniaturisation places constraints on the physical dimensions of the material components, which are often suppressed to the nanometer length scale. In materials of reduced dimensions, where the surface plays a major role, it then becomes a question of the impact of the surface on the magnetic properties.  

As the field of topological magnetic textures is strongly inspired and shaped by the observations made in bulk materials, there is great interest to create samples of reduced dimensions, such as thin films or nano-structured specimens, from those materials which display long-wavelength, topological spin textures in their bulk. Yet, growing evidence establishes that the properties of bulk materials prepared in reduced dimensions differ substantially from genuine bulk dimensions. Typical open questions may be illustrated nicely in the class of cubic chiral magnets, where the magnetic phase diagrams differ distinctly between bulk samples and epitaxial films \cite{2011_Yu_NatureMaterb,2017_Wiedemann_, PhysRevB.84.060404,PhysRevB.88.214420, doi:10.7566/JPSJ.84.104708,PhysRevLett.108.267201, Park2014, wolf2022}. Further, epitaxial films prepared from cubic chiral magnets present an interesting case where a determination of the ground state magnetic configuration is unresolved \cite{PhysRevB.86.144420,PhysRevB.88.214420,doi:10.7566/JPSJ.84.104708,2016_Kanazawa_PhysRevB,2018_Zheng_NatureNanotechc}. In addition, the surfaces of cubic chiral magnets appear to support an unexplained strong N\'{e}el twisting \cite{2018_Zhang_ProcNatlAcadSciUSAa}. Last but not least, even metastable surface configurations, known as chiral bobbers, may completely change the energetics of samples with reduced dimensions \cite{2019_Redies_PhysRevBc}. 

This illustrates the need for the detailed experimental determination of the magnetic structure of systems prepared with reduced dimensions. While real-space imaging techniques such as spin-polarized tunneling, magnetic force, or Lorentz transmission electron microscopy may provide detailed microscopic insights on local scales, their applicability for bulk materials, as well as extended films, is quite limited and their use rather demanding technically \cite{doi:10.1126/science.1234657,doi:10.1021/nl401687d,doi:10.1038/nphys2045,PhysRevLett.110.117202,PhysRevLett.112.059701}. Likewise, despite major advances in recent decades, X-ray scattering appears to be valuable in studies of magnetic structures with long-wavelength modulations under very specific conditions only \cite{PhysRevB.93.214420,PhysRevLett.112.167202}. 

In comparison, as a weakly interacting probe matching the length- and time-scales of long-wavelength magnetic modulations, neutron scattering has been key in determining the reciprocal-space characteristics of a vast range of materials. In particular, small-angle neutron scattering (SANS) has played a prominent role in allowing for the characterisation of structural, magnetic and chemical periodicities on the nanometer length scale \cite{RevModPhys.91.015004}. To enable high angular resolution on a SANS instrument, neutrons are strongly collimated over large distances (typically, 1 -- \SI{20}{m}) to minimise beam divergences, and background contributions are limited through the use of pin-hole apertures. Therefore, a major caveat of SANS is that measurements are largely intensity limited due to a restricted flux at the sample position. To improve the signal-to-background ratio, sample volumes on the order of 1 mm$^3$ are typically required.

Neutron scattering studies of systems of limited dimensions to date have been constrained strongly by the combination of tiny sample volumes and background scattering due to substrates or sample support structures. A key approach to avoid such scattering contributions is the use of reflection geometries, where the surface only contributes the scattering signal. This has resulted in major advances in specular and off-specular (polarized) neutron reflectometry, which profiles the depth evolution of magnetic moments in thin films and layered materials with sub-nanometer depth resolution $Q_z$ \cite{FELCHER1993137}. Yet, as a major drawback, neutron reflectometry requires the assumption of complex scattering profiles, where the uniqueness of a solution is not guaranteed. Moreover, the technique averages in-plane structure along the $y$-direction, due to relaxed beam collimation, which prevents $Q_y$ resolution in the plane of the film \cite{SAERBECK2014237}. As such, reflectometry is unable to provide sufficient angular resolution in both $Q_x$ and $Q_y$ required to fully resolve complex lateral correlations.

The need for direct imaging of the reciprocal-space properties of long-wavelength magnetic modulations in systems of reduced sample dimensions has reinvigorated the interest of how to optimize the signal-to-background ratio in small-angle neutron scattering. An obvious approach pursues the reduction of the background scattering as compared to conventional transmission geometries. A rather elegant configuration that meets this requirement may be achieved in grazing-incidence SANS (GI-SANS), where the neutrons form an evanescent wave in the top-most layer of the sample \cite{dosch}. However, the conditions for GI-SANS depend sensitively on material-specific details that cannot be met in many materials of great interest. It is nonetheless possible to improve the signal-to-background ratio considerably close to, but slightly above, the critical angle of reflection. This so-called near-surface SANS (NS-SANS) geometry allows to minimize scattering contributions by the substrate while providing essentially identical scattering patterns as in a transmission measurement, albeit with the addition of a specular reflection. 

Our paper is organized as follows. In Section \ref{geometries}, we present a brief introduction of the SANS geometries of basic interest, notably conventional transmission SANS, near-surface SANS, and grazing-incidence SANS. This is followed, in Section \ref{proof}, by an account of proof-of-concept measurements on the non-centrosymmetric cubic chiral magnet MnSi which exhibits a well-documented magnetic phase diagram \cite{science2009}. Our experimental results provide certainty that NS-SANS, with its modified scattering geometry compared to transmission SANS, provides a complete depiction of the nanoscale periodicities present within a magnetic sample. It showcases the applicability of NS-SANS for the study of nano-confined materials, which would otherwise possess insufficient scattering volumes to be measured in transmission. The experimental results motivate the prediction of the neutron scattering patterns arising from thin films of MnSi in Section \ref{prediction}, provided the same magnetically modulated phases would form as in bulk MnSi. Considering the Fourier transformation of the real-space magnetic configurations, we present a complete catalogue of scattering patterns arising from long-wavelength magnetic modulations for each orientation of the three-dimensional order with respect to the scattering plane. Our paper finishes in Section \ref{conclusion} with a brief account of the main conclusions.

\section{Description of SANS Geometries}
\label{geometries}

This section outlines the different scattering configurations discussed in this paper. We begin with a description of transmission SANS followed by near-surface and grazing-incidence SANS geometries. For simplicity, our discussion applies to a disk-shaped sample of surface area $A$ and edge thickness $t$. We assume that the footprint of the neutron beam is smaller than $A$ in transmission geometry, and smaller than the projected $A$ in the near-surface and grazing-incidence geometries. We adopt a sample-centric coordinate system where the $z$-axis is always taken as the direction parallel to the surface normal (i.e. axis parallel to $t$).

\subsection{Transmission SANS}

Figure \ref{fig:1}(a) displays the geometry of conventional SANS performed in transmission mode, where a tightly collimated beam of monochromatic neutrons are directed  at normal incidence to the sample surface $A$ lying in the ($x$, $y$) plane. The edge thickness $t$  of the sample is positioned parallel to the incident neutron beam which propagates along the $z$-axis of the sample. Rocking scans are obtained by rotating the sample about the vertical $\omega$ and horizontal $\phi$ instrument axes. The detector is situated in the ($x$, $y$) scattering plane, and is sensitive to scattering from lateral structures in the ($x$, $y$) sample plane.

In transmission SANS, incident neutrons scatter within the sample volume and propagate towards an area detector at a given scattering angle 2$\theta$. A scattering pattern of intensity $I(Q)$ is recorded on the detector as a function of the scattering vector
\begin{equation}
    Q = \frac{4\pi}{\lambda} \sin \left(\frac{2\theta}{2}\right)
    \label{eq:1}
\end{equation}
in units of inverse length, where $\lambda$ is the neutron wavelength. For a sample with short-range order, the resultant $I(Q)$ will form a ring of intensity centered about the direct beam location with a radius of $|Q|$. For a sample with long-range order, distinct Bragg peaks will arise at particular $Q$ locations on the detector, usually along high-symmetry or crystallographic directions. The position of the peaks discloses the size of the periodicities in real-space, where $d = 2\pi/Q$. In the SANS regime, bulk structures with typical real-space dimensions on the order of 1 -- 300 nanometers are probed.

It has long been appreciated that transmission SANS is not ideally suited to the study of samples of reduced dimensions, as the scattering volume is too small. As such, the use of transmission SANS to characterise materials of reduced dimensions, such as magnetic periodicities in thin films or microstructured samples, has not been convincingly reported on a single film comprised of a few magnetic repetitions only \cite{meynell}. In comparison, stacking several identically prepared multilayers on top of another proved sufficient to increase the SANS signal to an appreciable level above background \cite{PhysRevMaterials.3.081401,PhysRevMaterials.3.104406}. Nevertheless, studies of this nature require a large number of co-aligned films each supported by a substrate, where each substrate contributes to the scattering signal, impacting the overall data statistics. Furthermore, potential misalignment issues between adjacent films in the multilayer stack can result in an undesired smearing of diffraction peaks which can lead to data ambiguity.

\subsection{Near-Surface SANS}

Near-surface SANS (NS-SANS) provides a route to overcome the shortcomings of transmission SANS for extremely small sample volumes in the thin-film limit. Namely, NS-SANS is in many ways more powerful and universal, even if grazing-incidence scattering can be achieved. NS-SANS measurements are performed on SANS instruments using tightly collimated neutrons directed at grazing incidence to the sample at an angle greater than the critical angle of reflection \cite{hamilton2005}. 

The NS-SANS geometry capitalises on negligible refraction effects, which allows most of the beam to enter the sample to probe nanoscale periodicities present within the bulk of the sample volume. For NS-SANS, it is necessary to select samples with small neutron absorption cross-sections to aid the extended neutron path length in the material. An additional advantage of NS-SANS is that the sampling depth can be tuned by varying the angle of the incident neutron beam. For NS-SANS investigations on thin films, this implies it is possible to avoid substrate contributions which otherwise form a source of background noise in transmission SANS measurements. NS-SANS has been most widely employed in the field of soft matter for the study of polymer micelles \cite{WOLFF2007,Ruderer,kyrey}; however, to date, has not yet been comprehensively reported on for the study of magnetic nanostructures in low-dimensional condensed matter systems.

Figure \ref{fig:1}(b) displays the geometry of SANS performed in reflection mode, where a tightly collimated beam of monochromatic neutrons are directed at a grazing incidence angle $\alpha_i$ to the sample surface $A$ lying in the ($x$, $y$) plane. The edge thickness $t$ of the sample is positioned (almost) perpendicular to the incident neutron beam which propagates along the $x$-axis of the sample. Rocking scans are obtained by rotating the sample about the vertical $\omega$ and horizontal $\phi$ instrument axes. The detector is situated in the ($y$, $z$) scattering plane, and is sensitive to scattering from lateral and vertical structures in the ($y$, $z$) sample plane. A scattering horizon $h$ represents the sample-air interface \cite{hamilton1996}.

Near-surface SANS occurs in the reflection geometry exclusively for incidence angles $\alpha_i$ greater than or equal to the critical angle of reflection
\begin{equation}
    \alpha_c = \lambda \sqrt{\frac{Nb}{\pi}}
    \label{eq:2}
\end{equation}
of the sample, of a given density $N$ and scattering length $b$. At incidence angles greater than or equal to $\alpha_c$, the sample is fully illuminated and the bulk ordering of the sample can be probed due to the partial reflection and refraction of the neutron beam at the sample surface. The reflected component of the neutron beam undergoes specular reflection resulting in a specular peak on the detector at $\alpha_f$ = $\alpha_i$. The transmitted component of the neutron beam is refracted at the sample surface to undergo small-angle scattering within the bulk of the sample, and if re-scattered towards a sample surface, is transmitted through it with some probability. The transmitted neutrons propagate towards the detector to contribute scattering intensity at a given scattering angle 2$\theta$. Dynamical scattering in the NS-SANS geometry leads to an enhancement of the scattering intensity at $\alpha_f$ = $\alpha_c$, commonly referred to as the Yoneda peak \cite{Yoneda}. Contributions to the scattering pattern occur both above and below the scattering horizon $h$, where intensities recorded below $h$ are typically weaker due to greater beam absorption by the sample holder. 

For the example of NS-SANS scattering from a sample exhibiting long-range magnetic order, the position of the specular peak at $\alpha_f$ will be independent of temperature and field, and will only depend on the selected $\alpha_i$ and $\lambda$. The sharpness of the specular peak will depend on the roughness of the sample surface, and the intensity will be governed by the neutron absorption cross-section. In contrast, the position of the magnetic Bragg peaks arising at $Q$ from the internal magnetic order will be completely independent of $\alpha_i$, but will exhibit field and temperature dependencies. The range of the magnetic order (i.e. short-range or long-range) will influence the sharpness of the Bragg peaks. For an under-illuminated sample, the intensity of the peaks will increase with decreasing incidence angle (for $\alpha_i >\alpha_c$) due to increases in beam footprint.

\subsection{Grazing-Incidence SANS}

At incidence angles less than the critical angle ($\alpha_i <\alpha_c$) of the sample, a second scattering regime, known as grazing-incidence SANS \cite{Nouhi}, is encountered in the reflection geometry. For incidence angles less than $\alpha_c$ all incident neutrons are totally externally reflected from the sample surface and only an evanescent wave extends into the sample volume \cite{dosch}. The evanescent wave is exponentially damped within a few nanometers of the surface, where the probed volume is limited by the penetration depth of the neutrons
\begin{equation}
    D = \frac{\lambda}{\sqrt{2} \pi (l_i + l_f)}
    \label{eq:3}
\end{equation}
where
\begin{equation}
    l_{i,f} = \sqrt{{{\alpha_c}^2}-{\alpha_{{i,f}}^2} + \left[{\left({\alpha_{{i,f}}^2}-{\alpha_c^2}\right)^2 + \left(\frac{\lambda\mu}{2\pi}\right)^2}\right]^{1/2}}
    \label{eq:4}
\end{equation}
and $\mu$ is the attenuation coefficient.

GI-SANS is distinct from NS-SANS in that it is purely a surface sensitive measurement that is performed at incidence angles less than the critical angle of the material under study. GI-SANS is ideally suited to investigating the size and shape of nanostructures positioned on top of a surface, or lateral correlations or coherent ordering located within a few 100 nanometers of a surface \cite{pmb2012}. GI-SANS is treated within the framework of the distorted-wave Born approximation and is therefore equivalent to conventional reflectometry measurements \cite{FELCHER1993137}, but offers greater resolution in $Q_y$ through the use of a non-divergent beam.

\begin{figure}
\begin{centering}
\includegraphics[width=0.8\textwidth]{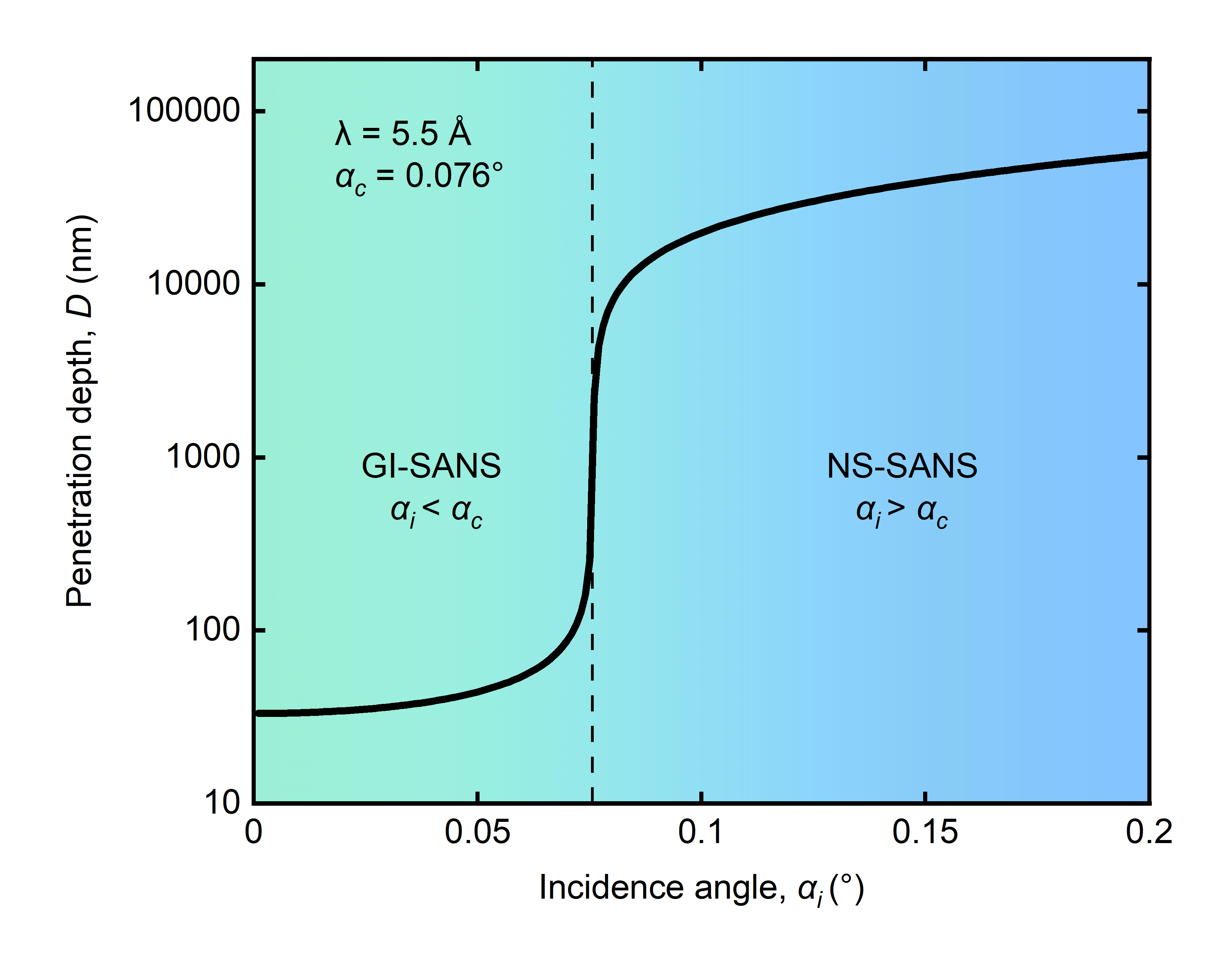}
\caption{Relationship between the penetration depth $D$ and the incidence angle $\alpha_i$ of the neutron beam. Results are calculated for MnSi at a neutron wavelength of \SI{5.5}{\text{\r{A}}}, resulting in a critical angle of $\alpha_c =$ \SI{0.076}{\degree}. For incidence angles less than the critical angle ($\alpha_i <\alpha_c$) surface sensitive GI-SANS takes place. For incidence angles greater than or equal to the critical angle ($\alpha_i \geq \alpha_c$) bulk sensitive NS-SANS takes place.}
\label{fig:2}
\end{centering}
\end{figure}

Optimising reflection-mode SANS measurements for either surface or bulk sensitivity is achieved by varying the angle of the incident neutron beam below and above the critical angle $\alpha_c$ of the material. Changes to the scattering pattern are observed as the incidence angle of the incoming beam increases and probes greater depths in the material. Figure \ref{fig:2} exemplifies the penetration depth of \SI{5.5}{\text{\r{A}}} neutrons reflected from the surface of bulk MnSi calculated from Equations \ref{eq:3}-\ref{eq:4}. At the fixed neutron wavelength, the critical angle of MnSi is $\alpha_c = $ \SI{0.076}{\degree}. In the GI-SANS regime where $\alpha_i <\alpha_c$, the probed volume is restricted by the neutron penetration depth that is limited to a few tens of nanometers from the sample surface. At increased incidence angles above the critical angle, the neutron penetration depth is greatly enhanced to more than 10$^4$ nanometers below the sample surface, and bulk sensitivity is achieved in the NS-SANS regime.

\section{Long-Wavelength Magnetic Modulations}
\label{proof}

This section describes the most common forms of long-wavelength magnetic modulations reported in the literature, and includes a description of helical, conical and skyrmion lattice orders. A class of materials which has been shown to support all the abovementioned forms of non-collinear magnetic order are the family of chiral magnets (e.g. MnSi, FeGe, Fe$_{1-x}$Co$_x$Si and Cu$_2$OSeO$_3$) which crystallise in the B20 cubic structure with $P$2$_1$3 space group symmetry \cite{bauer2016}. In the following sections, the cubic chiral magnet of MnSi is chosen as the working example as it has a well-documented phase diagram \cite{bauer2012}, a cubic lattice and exhibits non-collinear magnetic periodicities on the length scale applicable to small-angle scattering \cite{science2009}.

\subsection{Magnetic Phases of B20 MnSi}

The cubic chiral magnet of MnSi exhibits a non-centrosymmetric crystal structure which lacks inversion symmetry and enforces the Dzyaloshinskii-Moriya spin-orbit interaction. This particular interaction generates a small continuous twist between neighboring spins, resulting in long-period helical spin modulations.

The helical phase of MnSi stabilizes below a critical temperature of $T_c \sim$ \SI{29}{\kelvin} in zero applied magnetic field $B$. The helical phase forms in a multi-domain state where equally populated domains of helices propagate along the four equivalent \hkl<111> crystallographic easy-axes with a propagation length of \SI{180}{\text{\r{A}}}. In reciprocal space, each helical domain gives rise to a pair of satellite Bragg peaks located at $Q = \pm$ 0.035 \AA$^{-1}$ along the \hkl<111> directions. For an incident neutron beam parallel to the \hkl<110> crystallographic axis of MnSi, the scattering plane will be sensitive to four individual Bragg peaks corresponding to propagations along $\pm Q$ $||$ \hkl[111] and $\pm Q$ $||$ \hkl[-1-11]. 

In the presence of small applied magnetic fields $B$, the domain degeneracy of MnSi is lifted as the propagation vectors $Q$ start to cant towards $B$ whilst preserving their magnitude. At the characteristic field $B_{c1}$ the system transitions into the conical state where $Q$ is parallel to $B$. As a result, the conical phase gives rise to two Bragg peaks located at $\pm Q$ $||$ $B$ in reciprocal space. With increasing $B$ the canting angle of the conical order is continuously reduced and vanishes completely at $B_{c2}$ where the systems enters the field-polarised phase.

A topological skyrmion phase is stabilised by thermal fluctuations in a small phase pocket just below $T_c$ in intermediate fields. The skyrmion lattice phase of MnSi can be understood as the superposition of three helical propagations leading to a hexagonal arrangement of magnetic swirls in the plane perpendicular to $B$. This periodic arrangement of spins translates in reciprocal space to the appearance of six Bragg peaks located equidistant from the direct beam and each separated by 60 degrees. 

As a function of increasing temperature, the helimagnetic-to-paramagnetic transition at zero magnetic field displays characteristics consistent with a Brazovskii-transition \cite{PhysRevLett.110.177207}. In the paramagnetic regime, the Dzyaloshinsky-Moriya interactions generate a strong helical character, while the magnetic anisotropies are vanishingly small, resulting in an enhancement of the phase-space available for fluctuations such that the transition is driven first order. These strong helimagnetic fluctuations above $T_c$, which are referred to as the  fluctuation-disordered regime \cite{2013_Janoschek_PhysRevBa}, cause the resulting scattering intensity to spread out over a sphere in reciprocal space without the formation of distinct Bragg peaks.

\subsection{Equations of Long-Wavelength Magnetic Order}

In the simplest approximation, ignoring the effects of weak cubic magneto-crystalline anisotropies, the different forms of magnetic order described qualitatively above, may be accounted for in terms of the harmonic modulations described in the following.

\subsubsection{Helical Order}
In the helical phase, neighbouring spins uniformly rotate in the plane perpendicular to the propagation direction. Helical magnetic order is modelled by 
\begin{equation}
    \bm{m}(\bm{r}) = \left(\begin{matrix} \cos (\bm{k}_i\cdot\bm{r}\frac{2\pi}{\lambda})\\[0.2cm]  \sin(\bm{k}_i\cdot\bm{r}\frac{2\pi}{\lambda})\\[0.2cm]  0\end{matrix}\right)
    \label{eq:5}
\end{equation}where the propagation vector is parallel to the $z$-axis and $\lambda$ is the wavelength of the propagation.

\subsubsection{Conical Order}
In the conical phase, neighbouring spins uniformly rotate with an opening angle $\alpha$ to the propagation direction. Conical magnetic order is modelled by
\begin{equation}
    \bm{m}(\bm{r}) = \left(\begin{matrix} \cos (\bm{k}_i\cdot\bm{r}\frac{2\pi}{\lambda})\sin (\alpha)\\[0.2cm]  \sin(\bm{k}_i\cdot\bm{r}\frac{2\pi}{\lambda})\sin(\alpha) \\[0.2cm]  \cos(\alpha)\end{matrix}\right)
    \label{eq:6}
\end{equation}where the propagation vector is parallel to the $z$-axis and $\lambda$ is the wavelength of the propagation.

\subsubsection{Skyrmion Lattice Order}
In the skyrmion lattice phase, spins are described by the superposition of three helical propagations separated by 120 degrees in the plane perpendicular to $B$. The skyrmion lattice phase can be modelled by 
\begin{equation}
 \bm{m}(\bm{r})=\sum_{i=1,2,3}\left [\left\{\begin{matrix} 0\\0\\-1\end{matrix}\right\}\cos\left(\bm{k}_i\cdot\bm r \frac{2\pi}{\lambda}\right)
-\left(\left\{\begin{matrix}0\\0\\-1\end{matrix}\right\}\times \bm{k}_i\right)
\sin\left(\bm{k}_i\cdot\bm{r}\frac{2\pi}{\lambda}\right) \right ],
 \label{eq:7}
\end{equation}
where 
\begin{equation}
\bm{k}_i = \left\{\begin{matrix} \cos(\alpha_i)\sin(\pi/2)\\\sin(\alpha_i)\sin(\pi/2)\\\cos(\pi/2)\end{matrix}\right\}
 \label{eq:8}
\end{equation} and $\alpha_{1,2,3} = \SI{0}{\degree},\SI{120}{\degree}, \SI{240}{\degree}$ represent the angles under which the three helices are superposed in the ($x$, $y$) plane. 

\section{Experimental Results and Discussion}
\label{prediction}

The presentation of our results is organised as follows. We begin with a presentation of the neutron scattering data obtained on bulk MnSi in the transmission and near-surface SANS geometries. The excellent agreement obtained between the two geometries motivates detailed predictions of the scattering patterns expected from samples of reduced dimensions supporting the same magnetic structures. The scattering patterns are obtained by Fourier analysis of the real-space magnetisation densities in the helical, conical and skyrmion lattice phases of bulk MnSi.

As previously mentioned, MnSi was chosen as a working example as it is a prototypical non-collinear magnet exhibiting a well-documented phase diagram, cubic lattice and magnetic periodicities on the length scale applicable to small-angle scattering. The discussion and analysis presented in this section applies to the entire family of B20 cubic chiral magnets (e.g. FeGe, Fe$_{1-x}$Co$_x$Si and Cu$_2$OSeO$_3$).

\subsection{Experimental Neutron Scattering Patterns}

Measurements were performed on the SANS-1 beamline at the FRM II, Munich \cite{Mhlbauer2016TheNS}. All measurements were performed using a monochromatic neutron wavelength of \SI{5.5}{\text{\r{A}}}, a collimation distance of \SI{23}{m}, a detector distance of \SI{8.2}{m}, and a wavelength resolution of $\Delta \lambda / \lambda = 10 \%$. The source aperture was 50 x 25 mm$^2$, and a sample aperture of 1 x 15 mm$^2$ was employed to replicate the type of aperture required for investigations on nano-confined samples, and to allow for a valid comparison to our computational results. 

Data were collected for neutrons transmitted through or reflected from the surface of a high-quality MnSi single crystal prepared by the Czochralski method \cite{CzochralskiEinNV}. To facilitate a reflection plane in the NS-SANS geometry, the crystal was polished mirror-flat into the shape of a disk with approximate dimensions of 25 x 15 x 3 mm$^3$ corresponding to the crystallographic directions of  \hkl[110], \hkl[001], \hkl[1-10], respectively, which were determined by X-ray diffraction. In the following discussions, a sample-centric coordination system is employed where $\hat{x}$ $||$ \hkl[110], $\hat{y}$ $||$ \hkl[001] and $\hat{z}$ $||$ \hkl[1-10]. The orientation of the sample in conjunction with the symmetries of the magnetic structure forming in the cubic crystal environment resulted in identical scattering patterns in transmission and NS-SANS geometries.

\begin{figure}
\begin{centering}
\includegraphics[width=\textwidth]{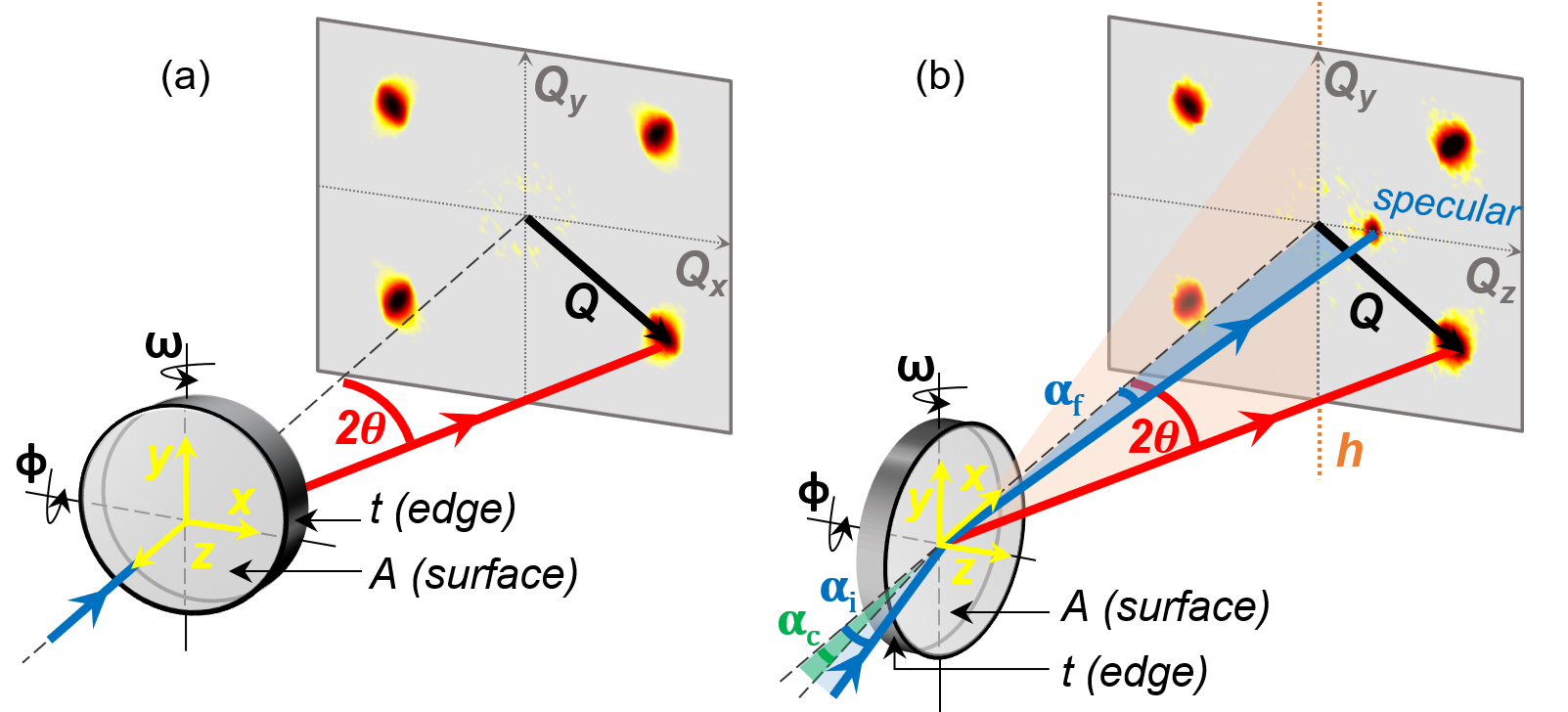}
\caption{Schematic depiction of the different SANS geometries. (a) Transmission SANS geometry where the neutron beam is oriented normal to the sample surface $A$ located in the ($x$, $y$) plane. The edge thickness $t$ of the sample is positioned parallel to the incident neutron beam which propagates along the $z$-axis of the sample. (b) Reflection SANS geometry where the neutron beam is directed at a shallow incidence angle $\alpha_i$ with respect to the sample surface $A$ located in the ($x$, $y$) plane. The edge thickness $t$ of the sample is positioned (almost) perpendicular to the incident neutron beam which propagates along the $x$-axis of the sample. For values of $\alpha_i$ greater than or equal to the material-specific critical angle of reflection $\alpha_c$, the configuration is known as near-surface SANS (NS-SANS). For values of $\alpha_i$ less than $\alpha_c$ an evanescent wave may form in the surface layers of the sample. This configuration is referred to as grazing-incidence SANS (GI-SANS). For each geometry, rocking scans are obtained by rotating the sample about the vertical $\omega$ and horizontal $\phi$ instrument axes.}
\label{fig:1}
\end{centering}
\end{figure}

In the transmission geometry, the sample was positioned on the beamline with the $y$-axis in the vertical direction and the $z$-axis parallel to the incident neutron beam, as depicted in Figure \ref{fig:1}(a). To transition to the NS-SANS geometry, the sample was rotated about its vertical axis until the $x$-axis of the sample was nearly parallel to the incident neutron beam, as depicted in Figure \ref{fig:1}(b). The surface of the sample was aligned to the incident neutron beam by maximising the intensity of the magnetic Bragg peaks. In the NS-SANS geometry, neutrons which entered the edge thickness of the sample -- rather than the polished surface -- were strongly absorbed (i.e. $\approx$ 1\% transmission over a 25 mm path length), and as a result near-surface contributions dominated the scattering signal. Rocking scans were performed by rotating the magnet and the sample together with respect to the neutron beam over $\pm$ \SI{4}{\degree} about the $\omega$ and $\phi$ instrument axes, as depicted in Figure \ref{fig:1}. The external magnetic field was directed perpendicular to the incident neutron beam in the conical phase, and parallel to the incident neutron beam in the skyrmion phase.  Due to cubic crystal symmetries neutrons always propagated along a \hkl<110> crystallographic axis of the sample.  

NS-SANS measurements were performed with the neutron beam at an incidence angle of $\alpha_i$ = \SI{0.3}{\degree} to the sample surface, well above the material- and wavelength-specific critical angle of $\alpha_c$ =  \SI{0.076}{\degree}. In accordance with Figure \ref{fig:2}, this resulted in a penetration depth of approximately 100 micrometers into the sample, allowing bulk periodicities to be probed. The choice of $\alpha_i$ also ensured good separation of the specular and magnetic Bragg peaks. To prevent saturation of the detector, a mask was placed over the direct beam which also masked the Yoneda peak arising at $Q_z$ = 0.003 \AA$^{-1}$ in the NS-SANS geometry. NS-SANS data were not corrected for refraction effects due to the negligible neutron scattering length density of MnSi \cite{WOLFF2007}.

\begin{figure}
\begin{centering}
\includegraphics[width=0.75\textwidth]{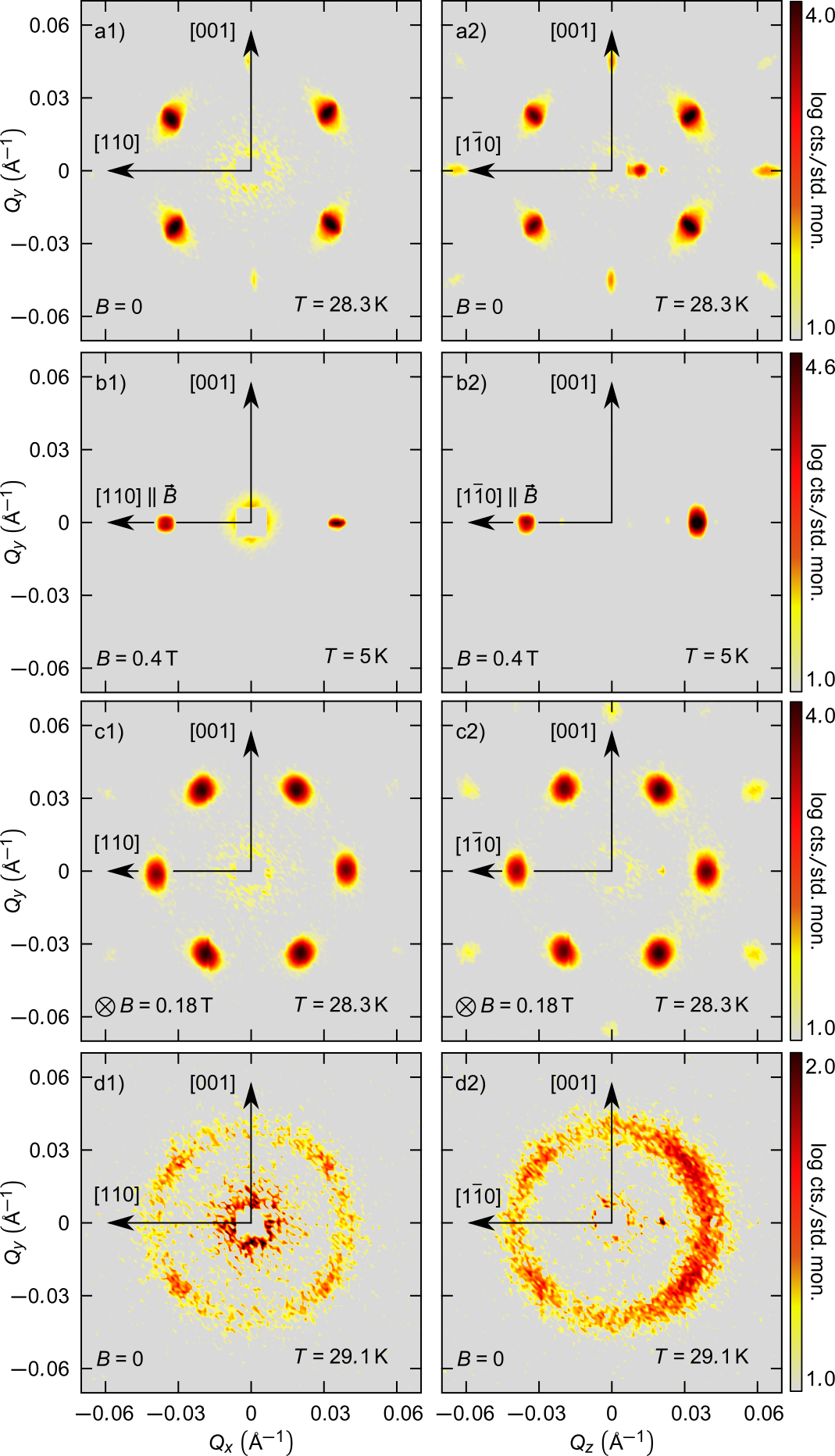}
\caption{Neutron scattering data. Scattering patterns obtained in the transmission SANS geometry (column 1) and in the NS-SANS geometry (column 2) in the (a) helical, (b) conical, (c) skyrmion, and (d) fluctuation-disordered phases of MnSi. Equivalent scattering patterns obtained between the two SANS geometries implies that the same magnetic periodicities are probed.}
\label{fig:3}
\end{centering}
\end{figure}

The experimental results are summarised in Figure \ref{fig:3}. Data represent rocking sums about the $\omega$ and $\phi$ axes. The transmission SANS data shown in Figures \ref{fig:3}(a1-c1) were obtained in the helical ($T$ = \SI{28.5}{\kelvin}, $B$ = \SI{0}{\tesla}), conical ($T$ = \SI{5}{K}, $B$ = \SI{0.4}{T}) and skyrmion phases ($T$ = \SI{28.3}{\kelvin}, $B$ = \SI{0.2}{T}) of MnSi. The scattering patterns exhibit the anticipated 4-fold, 2-fold and 6-fold arrangement of Bragg peaks that identify the respective real-space helical, conical and skyrmion spin configurations as shown in Figs.\,\ref{fig:3}(a1), (b1) and (c1), respectively. Likewise the scattering pattern measured at zero field in the fluctuation-disordered regime just above $T_c$, exhibits an essentially uniform ring of scattering with respect to the location of the direct beam, as shown in Fig.\,\ref{fig:3}(d1), as expected.

NS-SANS data obtained in reflection mode above the critical angle of MnSi at equivalent temperatures and fields is shown in Figures \ref{fig:3}(a2-d2). For each phase, the width of the Bragg peaks and the multiplicity and magnitude of the scattering vectors obtained in the NS-SANS geometry are equivalent to the transmission SANS data. This shows that regardless of the sample geometry with respect to the instrument geometry, the same magnetic periodicities are probed and imaged. Bragg peaks appearing below the scattering horizon (for $Q_z <$ 0) in NS-SANS have slightly reduced absolute intensities due to neutron absorption by the sample holder. Double scattering can be observed in some of the NS-SANS data, which was previously distinguished from second-order scattering in an earlier work \cite{adams2011}. The intensity of the double scattering likely arises as a result of the extended path length of neutrons in the sample due to the grazing-incidence geometry, combined with the magnetic mosaicity of the sample.

\begin{figure}
\begin{center}
\includegraphics[width=0.75\textwidth]{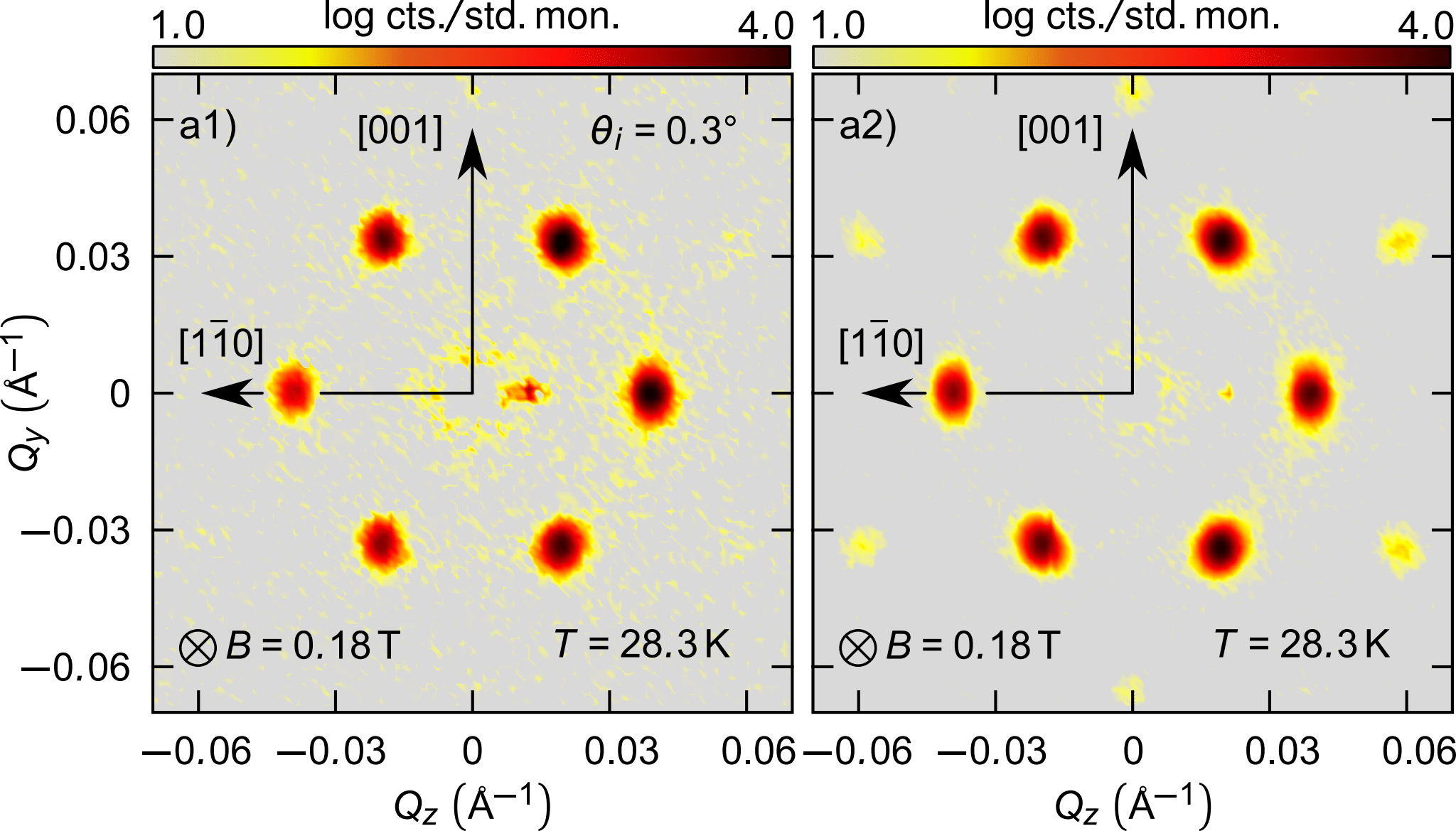}
\caption{Neutron scattering data. Scattering patterns obtained in the skyrmion phase of MnSi in the NS-SANS geometry. For data obtained (a1) at $\alpha_i$ = \SI{0.3}{\degree} without rocking, and (a2) as a rocking sum about the $\omega$ and $\phi$ axes.}
\label{fig:4}
\end{center}
\end{figure}

The scattering patterns of Figure \ref{fig:3} are plotted as rocking sums about the $\omega$ and $\phi$ axes, and as a result the specular reflection cannot be viewed in most cases. Figure \ref{fig:4} compares the scattering patterns obtained in the skyrmion phase of MnSi in the NS-SANS geometry at $\alpha_i$ = \SI{0.3}{\degree} without rocking in Fig. \ref{fig:4}(a1) and as a rocking sum about the $\omega$ and $\phi$ axes in Fig. \ref{fig:4}(a2). The NS-SANS data obtained without rocking is comprised of a 6-fold arrangement of Bragg peaks in addition to the specular reflection at $Q_z =$ 0.01 \AA$^{-1}$.

\subsection{Calculated Neutron Scattering Patterns of a Thin Film}

SANS patterns arising from the most common types of non-collinear magnetic modulations were calculated by taking the Fourier transformation of the real-space magnetic orders described in Equations \ref{eq:5}-\ref{eq:8}. The catalogue of results obtained in these calculations aim to provide insights as to what would be expected in NS-SANS studies, if the same magnetic modulations would be present as in genuine bulk samples.

Calculations were performed for a single-domain sample of MnSi exhibiting a bulk helical wavelength of \SI{180}{\text{\r{A}}}. The sample thickness was restricted to \SI{1000}{\text{\r{A}}} along the $z$-axis (infinite dimensions in both $x$ and $y$) to replicate the dimensions of a typical film in the thick-film limit. The effects of demagnetizing fields were ignored in the calculations. Fourier transformations of the magnetic order were convoluted with Gaussian distribution functions to simulate the effects of instrumental resolution. To avoid spectral leakage, a three-dimensional Blackman window function was applied to each magnetisation distribution before the Fourier transformation was calculated \cite{blackman1958}.
 
\begin{figure} 
\begin{centering}
 \advance\leftskip-4cm
\includegraphics[width=1.6\textwidth]{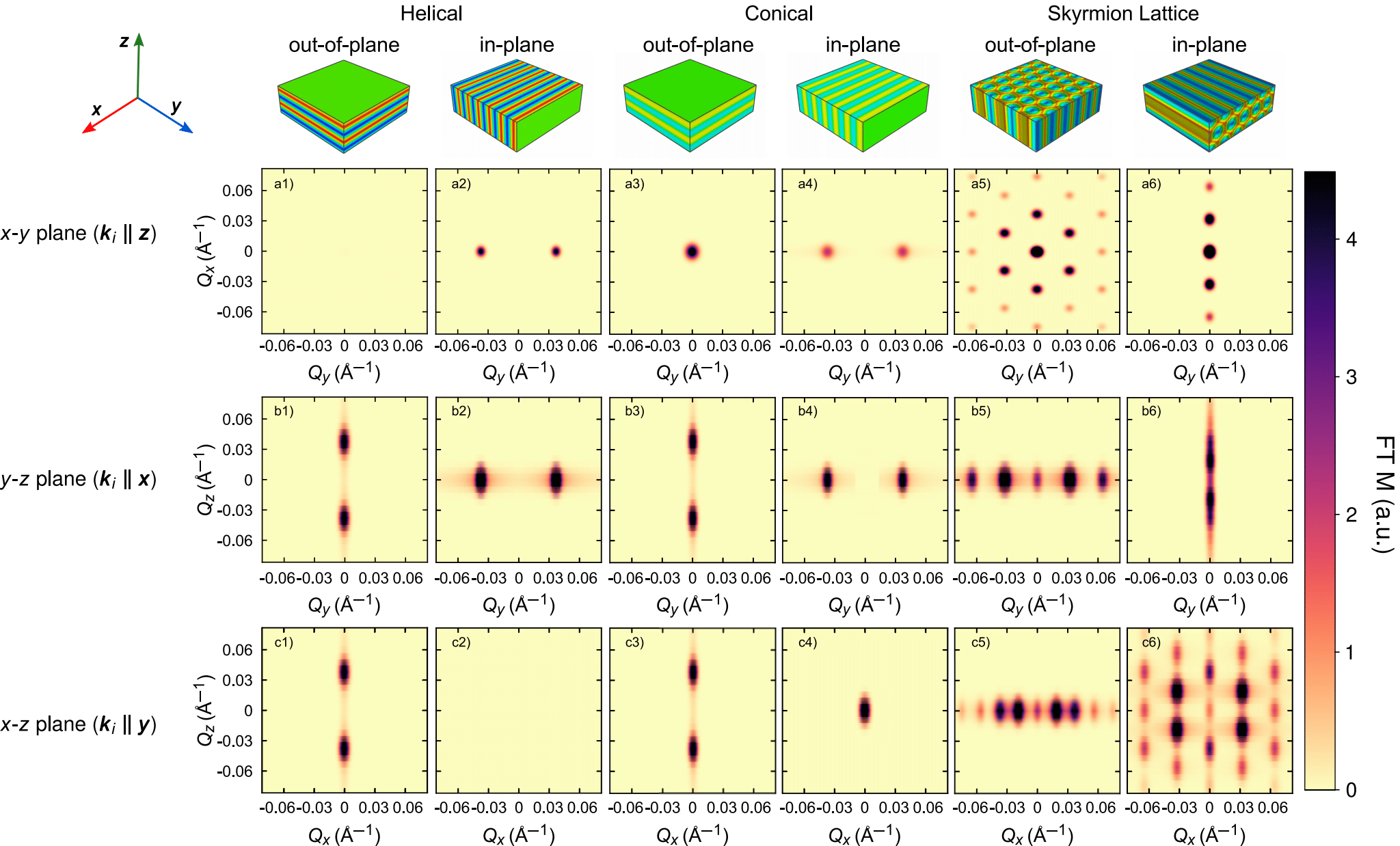}
\caption{Fourier transformations of the real-space magnetisation densities of helical, conical and skyrmion magnetic orders. Calculations were performed for a single-domain sample of MnSi exhibiting a bulk helical wavelength of \SI{180}{\text{\r{A}}} and a thickness of \SI{1000}{\text{\r{A}}}. The catalogue of results shown here serves as a point of reference for SANS studies of thin films in near-surface geometry.}
\label{fig:5}
\end{centering}
\end{figure}

The results of the calculations are summarised in Figure \ref{fig:5}. Calculations were performed for the ideal, real-space magnetisation configurations presented in the top row, comprising of single-domain helical, conical and skyrmion lattice orders. For completeness, both in-plane and out-of-plane propagations are considered. Depending on the direction of the incident wavevector $\bar{k_i}$ with respect to the three sample axes, each magnetic order can give rise to three distinct scattering patterns. 

The three possible scattering orientations, where $\bar{k_i}$ is either directed along the $z$-axis, $x$-axis or $y$-axis of the sample are shown in rows (a), (b) and (c) of Fig.\,\ref{fig:5}, respectively. In each case, the resulting scattering pattern will be sensitive to periodic modulations in the plane of the sample that is perpendicular to the incident $\bar{k_i}$. This implies that for the case of an out-of-plane propagating helix with $\bar{k_i} || \hat{z}$ as in Fig.\,\ref{fig:5}(a1) where the magnetisation is homogeneous in the ($x$, $y$) sample plane, no magnetic scattering will be observed on a detector lying in the ($x$, $y$) scattering plane. In contrast, as shown in Fig.\,\ref{fig:5}(a2), for $\bar{k_i}|| \hat{z}$ applied to an in-plane propagating helix exhibiting a periodicity along the samples' $y$-axis, the scattering pattern will exhibit 2-fold Bragg peaks at $\pm Q_y$ with a magnitude corresponding to the real-space wavelength of the helical periodicity.

The scattering patterns arising from helical and conical propagations share many commonalities for corresponding directions of $\bar{k_i}$ with respect to the sample axes, as plotted across columns 1-4 in Figure \ref{fig:5}. Comparison of in-plane helical and in-plane conical order for $\bar{k_i}|| \hat{z}$, in Figs.\,\ref{fig:5}(a2) and (a4), reveals qualitatively similar scattering patterns which display Bragg peaks at $\pm Q_y$ for magnetic propagations along the $y$-axis of the sample. Similarly, out-of-plane helical and out-of-plane conical order for $\bar{k_i}|| \hat{x}$ in Figs.\,\ref{fig:5}(b1) and (b3), both display Bragg peaks at $\pm Q_z$ due to the presence of one-dimensional magnetic propagations along the $z$-axes of the sample.

There are certain occurrences where conical magnetic order gives rise to an additional magnetic Bragg peak at zero scattering vector. The intensity at zero scattering vector can be observed for both out-of-plane  and in-plane propagating conical orders in Figs.\,\ref{fig:5}(a3) and (c4), respectively, which is otherwise absent from the corresponding helical order scattering patterns. Furthermore, the conical order Bragg peaks are often reduced in intensity compared to the helical order Bragg peaks. Both features are a consequence of the opening angle $\alpha$ of spins in the conical state, as described in Equation \ref{eq:6}. The opening angle results in a constant component of magnetisation perpendicular to the scattering vector, which acts like a periodicity of infinite wavelength giving rise to magnetic scattering at zero scattering vector. 

In any physical SANS experiment, the direct beam will be masked to avoid saturating the detector and hence any Bragg peaks that arise at zero scattering vector due to conical magnetic order will not be observed in reality. As a result, it can be difficult to distinguish between helical and conical magnetic orders using SANS, and hence it is often beneficial to decipher between these phases by other experimental means, such as magnetometry, AC susceptibility or ferromagnetic resonance.

An archetypal 6-fold scattering pattern is observed for the out-of-plane skyrmion lattice for $\bar{k_i}|| \hat{z}$, as shown in Fig.\,\ref{fig:5}(a5). Higher order reflections are observed in addition, as a consequence of the harmonic equations used to prepare the magnetic state. The remaining scattering orientations for an out-of-plane skyrmion lattice, shown in Fig.\,\ref{fig:5}(b5) and (c5) where $\bar{k_i}$ is either oriented along the $x$-axis or $y$-axis of the sample, will be sensitive to periodicities between the skyrmion tubes which are oriented vertically along the $z$-axis of the sample. As the only magnetic periodicities exist in the planes perpendicular to the skyrmion tubes, the scattering patterns will only exhibit scattering intensities along $Q_x$ or $Q_y$ (depending on the scattering geometry used) without any scattering along $Q_z$. Furthermore, due to the hexagonal arrangement of the skyrmion lattice, different periodicities will be generated along $Q_x$ and $Q_y$ consistent with the periodicities observed in Fig.\,\ref{fig:5}(a5) for $\bar{k_i}|| \hat{z}$.

The influence of reduced sample dimension on the scattering patterns is highlighted in the comparison of Figs.\,\ref{fig:5}(a5) and (c6). In each case, skyrmion tubes are oriented along $\bar{k_i}$ and the three helical propagations superposed in the plane perpendicular to $\bar{k_i}$ are responsible for the resulting scattering pattern. For out-of-plane skyrmions with $\bar{k_i}|| \hat{z}$ in Fig.\,\ref{fig:5}(a5), a 6-fold scattering pattern consisting of sharp Bragg peaks is observed due to the infinite skyrmion lattice structure located in the ($x$, $y$) sample plane. In comparison, a lower resolution 6-fold scattering pattern with enhanced secondary scattering is observed for in-plane skyrmions with $\bar{k_i}|| \hat{y}$ in Fig.\,\ref{fig:5}(c6). In the later case, the extent of the skyrmion lattice is restricted along the $z$-dimension resulting in a reduced $Q_z$ resolution and a smearing of the Bragg peaks compared to Fig.\,\ref{fig:5}(a5). The smearing of Bragg peaks is therefore anticipated in the scattering patterns of samples with reduced dimensions. As a result, these experimental considerations should be taken into account whenever a scattering pattern of a thin film or nano-structured specimen is to be interpreted.

\section{Conclusions}
\label{conclusion}

In conclusion we reported considerations on the geometries of SANS suitable for studies of long-wavelength magnetic modulations in samples of reduced dimensions. In view that transmission SANS is subject to a prohibitively large background signal and the conditions required for generic grazing-incidence SANS are difficult to satisfy, we consider the potential of near-surface SANS to provide a complete depiction of nanoscale periodicities present within nano-confined magnetic samples. Performing proof-of-concept measurements we show that the scattering patterns observed for a bulk sample measured in transmission geometry, and a polished surface of the same bulk crystal measured in the near-surface geometry are essentially identical. This motivates the calculation of the scattering patterns to be expected of MnSi thin film samples, if the magnetic structures are the same as in genuine bulk samples. While there are certain subtle differences, none of the generated scattering patterns match the properties reported in the literature for epitaxial films, asking for further exploration.

\section{Acknowledgements}

We wish to thank P. B\"oni, S. Mayer, A. Book, T. Meier, and S. M\"uhlbauer for discussions and support with the experiments. We also wish to thank the staff at the Heinz Maier-Leibnitz Zentrum (MLZ) for support. This work has been funded by the Deutsche Forschungsgemeinschaft (DFG, German Research Foundation) under TRR80 (From Electronic Correlations to Functionality, Project No. 107745057, Project F7), the priority program SPP 2137 (Skyrmionics) under grant PF393/19 (project-id 403191981), and the excellence cluster MCQST under Germany's Excellence Strategy EXC-2111 (Project No. 390814868). Financial support by the European Research Council (ERC) through Advanced Grants No. 291079 (TOPFIT) and No. 788031 (ExQuiSid) is gratefully acknowledged.

\pagebreak

\bibliographystyle{plain}
\bibliography{apssamp}

\pagebreak

\end{document}